\documentclass[preprint,12pt]{elsarticle}

\usepackage{tabularx}
\usepackage{booktabs}
\usepackage{multirow}
\usepackage{array}
\usepackage[most]{tcolorbox}
\usepackage{float}
\usepackage{placeins}

\journal{Information and Software Technology}
\begin{document}
\begin{frontmatter}



\title{A Systematic Literature Review on Logging Smell Detection}


\author [a] {Nora Madi\corref{corresponding}}
%% Author name
\ead{nmadi@ksu.edu.sa}

\author [a] {Manal Binkhonain}
\ead {mbinkhonain@ksu.edu.sa}

\cortext[corresponding]{Corresponding author}

\affiliation [a]{organization={Department of Software Engineering, College of Computer and Information Sciences, King Saud University},%Department and Organization
            addressline={}, 
            city={ Riyadh},
            postcode={12372}, 
            state={},
            country={Saudi Arabia}}

%% Abstract
\begin{abstract}
\textbf{Context:} Logging is an important part of software development that helps developers monitor systems, understand behavior, and fix problems. But when logging is done poorly, it can introduce logging smells, which are defects that reduce the usefulness of logs or even make them problematic. \\
\textbf{Objective:} This study looks at how logging smells are currently detected. The goal is to better understand the existing research on automatic detection techniques, datasets, and evaluation methods. \\
\textbf{Method:} We conducted a systematic literature review (SLR) of 21 studies focused on detecting logging smells. In this review, we define key logging-related terms, identify and map the types of smells to an existing taxonomy, and examine the detection techniques, datasets, and evaluation strategies used across the studies. \\
\textbf{Results:} We found that the research is still scattered and inconsistent. For example, there is no common benchmark or standardized approach for evaluating results, making it difficult to compare studies. In addition, we observe inconsistencies in the way log smells are addressed, as studies differ in the types and number of smells they target. \\
\textbf{Conclusion:} There is still room for improvement in how logging smells are studied and detected. We point out several challenges and suggest future directions, such as developing better tools, using large language models (LLMs), and building more standardized datasets for evaluation.
\end{abstract}



%% Keywords

\end{frontmatter}



\section{Introduction}
\label{sec1}
%% Labels are used to cross-reference an item using \ref command.
Software development is a complex and demanding process that requires developers to continuously understand and manage system behavior. Logging is widely used to support this need, especially during debugging, monitoring, and maintenance \cite{geng2024large}. Well-written logs provide developers with valuable insight into how systems are operating. However, logging is not always implemented effectively, causing log issues or log smells that can lead to confusion, missing information, degraded system performance, or reduced accuracy in downstream analytics tasks such as log parsing \cite{sedki2024decoding}. 

In recent years, numerous studies have attempted to automatically detect logging issues. Despite this progress, the existing literature lacks a systematic understanding of the methods and techniques employed. To address this gap, this study explores which types of logging smells are commonly targeted and what techniques are used to detect them. We conduct an SLR of 21 studies on logging smell detection to examine the types of issues adequately or inadequately addressed, the detection techniques applied, and evaluation strategies adopted. We also highlight common challenges and propose directions for future research.

Our analysis leads to two key observations. First, different studies focus on different types and numbers of issues, use different terms, follow different evaluation methods, and rely on various datasets. This suggests a lack of standardization in the field, which makes it difficult to compare results, reproduce findings, or build on existing work in a consistent manner. Second, while we are witnessing rapid advancements in the capabilities of LLMs in areas such as code comment generation \cite{xu2024unilog} and log statement generation \cite{geng2024large}, it is still rare to see these techniques applied in the context of log issue detection. However, detecting issues in generated content is just as important as generating it, as flawed or low-quality logs can introduce risks throughout the software development life cycle. 

The main contributions of this paper can be summarized as follows:
\begin{itemize}
  \item It presents a comprehensive understanding of logging issue detection research, based on an SLR of 21 studies.
  \item It maps existing software log smells to a structured taxonomy, helping reveal underexplored types that present opportunities for future research.
  \item It analyzes the variety of assessment methods and datasets used, showing inconsistencies that hinder cross-study comparisons.
  \item Based on the synthesis of reviewed studies and identified challenges, it suggests several directions for future research, including LLM-based detection, benchmarking, and tooling integration.
\end{itemize}

The remainder of this paper is organized as follows. Section 2 introduces the background and defines key terminology related to logging. Section 3 reviews related work on log analysis and log issue detection. Section 4 describes our methodology, including the research questions, data collection, and analysis approach. Section 5 presents the results of the review. Section 6 discusses threats to validity and Section 7 concludes the paper.

\section{Background}
\label{sec2}
This section explains software logging, providing a background of relevant terminology and an overview of an existing logging smell taxonomy.
\subsection{Software Logging}
\label{subsec2.1}
In this section, we aim to present and define the various terminology in the context of log-related research, along with alternative expressions that refer to the same or closely related concepts using previous reviews as  references (i.e., \cite{chen2021survey}, \cite{he2021survey}, \cite{gholamian2021comprehensive}, \cite{gu2022logging}, \cite{batoun2024literature}, and \cite{saarimaki2024taxonomy}).
\begin{itemize}
  \item \textbf{Logging Code:}
 
  The part of the source code responsible for generating log output. It includes log statements, logging configurations, and any surrounding control logic (e.g., conditions or guards) that determines when and how logs are produced. Logging code is inserted during the instrumentation phase and plays a key role in the quality of runtime monitoring and analysis.
  \item \textbf{Log Statement:} 
  
  A line of code written by developers that outputs runtime information during program execution. Also referred to as \textit{log printing (or generating) statement} or \textit{logging statement}. 
  
  The log statement includes a verbosity level (e.g., INFO or DEBUG), a static text message, and dynamic variables that capture runtime values.

  \item \textbf{Log Message:}
  
  The actual output of a log statement during execution. It includes a static description and dynamic variable values. Synonymous with \textit{log entry}, \textit{log record}, or \textit{log event}.
  \item \textbf{Log Level:}
  
  A classification that indicates the severity or importance of a log message, such as DEBUG, INFO, WARN, ERROR, or FATAL. It helps manage log verbosity and filtering.
  \item \textbf{Log:}
  
  A general term often used to refer to either individual log messages or entire \textit{log files} containing system runtime data. In literature, logs may also be referred to as \textit{execution logs}, \textit{log records}, or \textit{log events}.
  \item \textbf{Log Configuration:}
  
  The set of defined rules that tells the logging system how, where, and what to log (e.g., log level, output destination, and message format), thereby controlling the overall behavior of logging.
  \item \textbf{Logging:}
  
  The overall process of recording events during software execution by inserting log statements into the code. Logging involves the developer’s decision-making on \textit{what} and \textit{where} to log, followed by the instrumentation (i.e., inserting log statements in the source code), the storage of generated logs, and the subsequent analysis and management of log data. It supports various software engineering tasks, including debugging, monitoring, and fault diagnosis.
  \item \textbf{Logging Practice:}
  
  The methods and strategies developers use to manage logging throughout the software lifecycle, from insertion and configuration to storage and preprocessing for analysis.
  \item \textbf{Logging Issue:}
  
  A problem in how logging is implemented within software systems. Such issues can stem from various causes, including improper message content, configuration flaws, inadequate logging strategies, and log smells such as incorrect log levels or missing messages. These issues may lead to consequences such as misleading information, performance overhead, or unintended side effects during system execution.
  \item \textbf{Logging Smell:}
  
  A specific type of logging issue that may indicate poor logging practices and potentially lead to negative consequences such as reduced readability, maintainability, or effectiveness of log data, similar to code smells in source code. Examples include incorrect log levels, missing identifiers, or poor message formatting.
\end{itemize} 
Accordingly, in this study, we consider logging smells as a subset of logging issues and use the two terms interchangeably.
\subsection{Software Log Smell Taxonomy}
\label{subsec2.2}
Saarimäki et al. \cite{saarimaki2024taxonomy} present a systematic study about software log smells, by exploring bad practices and issues in software logging. The authors collected many logging issues from existing studies and created a taxonomy of log issues, categorized into causes, log smells, and consequences. This is the only log smell taxonomy that we know of that specifically focuses on comprehensively categorizing and addressing issues related to software logging quality. Therefore, we use it to provide a way to align the issues in the primary papers of this SLR, helping provide structure and consistency and offering a clearer basis for drawing meaningful conclusions.

This taxonomy of log smells includes ten smells with related facets that negatively impact logging quality. A facet is a specific variation of a log smell, detailing how that smell appears in practice (e.g, missing, inconsistent, redundant). The smells are:
\begin{itemize}
  \item \textbf{Format turmoil:} inconsistent or incomplete log formats that make logs hard to interpret.
  \item \textbf{Undercover identifier:} missing component or thread identifiers, which affect traceability.
  \item \textbf{Mercurial logging level:} incorrect or inconsistent severity levels, leading to confusion.
  \item \textbf{Deceptive variable:} misleading or malformed variable outputs that distort log information.
  \item \textbf{Variable on the edge:} nullable or uninitialized variables that lead to unreliable logs.
  \item \textbf{Message madness:} typos, duplication, or ambiguous messages that reduce clarity.
  \item \textbf{Logging lost in the wind:} missing or scarce log entries due to improper logging statements.
  \item \textbf{Landfill logs:} overly verbose or redundant logs that clutter log files.
  \item \textbf{Sleeping guards:} issues with logging guards that control whether logging code executes.
  \item \textbf{Skeleton in the closet:} problems in the code that generates the logs.
\end{itemize}
These smells often stem from causes (CA) such as a lack of logging guidelines, developer inexperience, insufficient tooling, or poor maintenance. They can lead to consequences (CO) including information leakage, incorrect event ordering, performance degradation, or changes in program behavior caused by logging.

Saarimäki et al. \cite{saarimaki2024taxonomy} found that some log smells, such as incorrect logging levels, missing logging guards, and missing messages, are well supported by detection and repair tools. In contrast, other smells like "Format turmoil" lack adequate tool support, highlighting gaps that need further research and development. Moreover, the results indicate that current tools often address only specific aspects of logging, with many focusing on a single type of issue. This leads to a fragmented approach that may require developers to rely on multiple tools to manage log quality effectively. 

These findings underscore the need to develop more comprehensive tools capable of addressing a broader range of log smells to improve overall logging quality.

\section{Related Work}
\label{sec3}
This section discusses existing literature reviews related to software logging quality. For example, Chen and Jiang \cite{chen2021survey} provide an overview of software logging research by reviewing 69 papers published between 1997 and 2019, 19 of which focus on logging quality. They identify nine main challenges, grouped into four categories: usability, diagnosability, logging code quality, and security compliance. Challenges related to logging code quality include clarity, maintainability, and consistency, highlighting the importance of writing logs that are understandable, kept up-to-date, and stylistically uniform across the codebase. The authors emphasize that the usefulness of log management tools depends on the quality of the logging, since high-quality logs help with tasks like monitoring and failure diagnosis. However, they also highlight that current techniques can detect only a small number of the issues in logging code—an observation that aligns with our own findings. Therefore, they suggest that it is important to develop benchmarks that can help compare different techniques more effectively. 

He et al.\cite{he2021survey} also address logging challenges as part of their study by reviewing papers published between 1997 and 2020. They survey 158 papers, 25 of which focus on logging. The challenges are categorized into three aspects: diagnosability, maintenance, and performance. The authors link the maintenance of logging code to the identification and characterization of anti-patterns, defined as poor coding practices in log statements that reduce log quality and increase maintenance effort. They note that empirical studies have shown these anti-patterns to be positively correlated with post-release defects, suggesting that developers should give higher priority to maintaining code with extensive logging. Although, this paper essentially focuses on automated log analysis such as log parsing, anomaly detection, and failure prediction. 

Similarly, Gholamian and Ward's survey, \cite{gholamian2021comprehensive} provides a taxonomy that categorizes current logging research into twelve topics, including logging practices and logging issues as subcategories under mining source code. Within these particular subcategories, they review 14 primary studies published between 2010 and 2021.  According to the authors, several studies have reported common logging issues such as anti-patterns, verbosity level problems, missing variables, and inconsistent static text. Other problems include duplicate logs and missing or inappropriate logging statements in important parts of the code. Detecting these logging problems can help developers make improvements to their log statements and increase log quality. Hence, the study points out gaps and opportunities for future research, including the development of advanced automated approaches for detecting inappropriate logging and the need for golden quality logging statements for benchmarking. These identified needs are also in line with our own observations, suggesting that there is still room for improvement and further research in this area.

Gu et al. \cite{gu2022logging} conducted a systematic mapping study (SMS) of 56 studies, categorizing them based on their focus: why-to-log, where-to-log, what-to-log, and how-well-to-log. The findings show that most research has focused on where-to-log and what-to-log, while the equally important aspects of why-to-log and how-well-to-log have received less attention. In addition, although what-to-log has been studied, the same log-related issues, such as those related to log levels and messages, are frequently explored, suggesting that there is still room for improvement and greater diversification in research topics. This observation aligns with our own findings, as we also notice that much of the reviewed work continues to focus on similar types of issues, leaving other important aspects underexplored.

A more recent study by Batoun et al. \cite{batoun2024literature} presents an SLR on software logging practices, analyzing 204 research studies related to instrumentation, storage, and analysis. Among these, they review 7 studies focused on log statement quality. According to Batoun et al., some prior work in this area has concentrated on analyzing log statements to detect logging issues, while other studies have addressed security-related aspects, such as logging vulnerabilities and techniques for protecting log data. Additionally, they conduct a qualitative study analyzing 149 StackOverflow logging questions using topic modeling and manual classification to uncover developers' common challenges and concerns in software logging, revealing gaps between research solutions and practical needs.

Each of the previous surveys explores general challenges and solutions related to logging quality as part of their work. In contrast, our study zooms in on the topic of log smell detection by focusing on different log issues and smells and the techniques used for detecting or assessing them, which helps reveal potential future research directions by providing a clearer understanding of the current state of research in this area. Our review also includes more recent studies that were not covered in previous work, quantitative analysis to provide an overview of different aspects of current log smell detection research, and an accumulation of reported challenges and future work from the reviewed papers, which we manually analyzed and enriched to highlight open gaps and opportunities for future research.

Table~\ref{tab:related_works} summarizes these studies alongside our own, which differs by providing a more focused investigation, whereas the other studies address only certain aspects of logging quality.

\FloatBarrier

\begin{table}[H]
\centering
\caption{Summary of related studies: title, year, focus, review method, coverage period, primary studies, and relevant subset counting primary studies that specifically address logging quality.}
\resizebox{\textwidth}{!}{%
\label{tab:related_works}
\renewcommand{\arraystretch}{1.3}
\begin{tabular}{|p{5cm}|c|p{5cm}|c|c|c|p{3.2cm}|c|}
\hline
\multicolumn{1}{|c|}{\textbf{Title}} & \multicolumn{1}{|c|}{\textbf{Year}} & \multicolumn{1}{|c|}{\textbf{Focus}} & \multicolumn{1}{|c|}{\textbf{Review Method}} & \multicolumn{1}{|c|}{\textbf{Period Span}} & \multicolumn{1}{|c|}{\textbf{Primary Papers}} & \multicolumn{1}{|c|}{\textbf{Relevant Subset}} & \multicolumn{1}{|c|}{\textbf{Ref.}} \\
\hline
\multicolumn{8}{|c|}{\textbf{Logs}} \\
\hline
A Survey of Software Log Instrumentation &
2021 &
Log instrumentation (approach, integration, composition challenges) &
SLR &
1997--2019 &
69 &
19 studies on logging quality &
\cite{chen2021survey} \\
\hline
A Survey on Automated Log Analysis for Reliability Engineering &
2021 &
Automated log analysis: logging, parsing, compression, mining &
SLR &
1997--2020 &
158 &
25 studies on constructing logging statements &
\cite{he2021survey} \\
\hline
A Comprehensive Survey of Logging in Software: From Logging Statements Automation to Log Mining and Analysis &
2022 &
Logging practices, mining, automation, and monitoring techniques &
SLR &
2010--2021 &
112 &
14 studies on logging practices and issues &
\cite{gholamian2021comprehensive} \\
\hline
Logging Practices in Software Engineering: A Systematic Mapping Study &
2023 &
Logging practices analyzed by Why, Where, What, and How well &
SMS &
2000--2022 &
56 &
17 studies on How well is the logging &
\cite{gu2022logging} \\
\hline
A Literature Review and Existing Challenges on Software Logging Practices &
2024 &
Logging practices across instrumentation, storage, analysis; gaps between research and practice &
SLR &
2013--2023 &
204 &
7 studies on log statement quality &
\cite{batoun2024literature} \\
\hline
\multicolumn{8}{|c|}{\textbf{This Study}} \\
\hline
A Systematic Literature Review on Logging Smell Detection &
-- &
Logging issues and smells, and detection or assessment techniques &
SLR &
Up to 2025 &
21 &
21 on detecting logging issues &
-- \\
\hline
\end{tabular}
}
\end{table}

\section{Methodology}
\label{sec4}

\begin{figure}[h]
\centering
\includegraphics[width=0.8\textwidth]{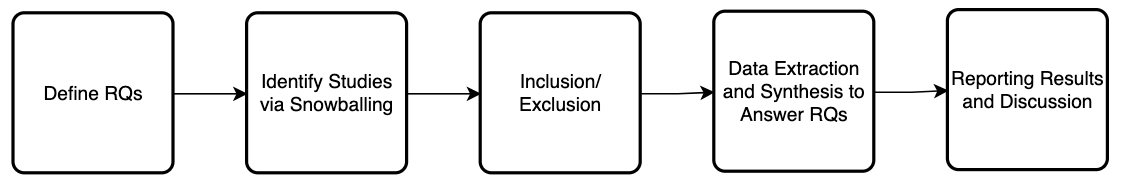}
\caption{Overview of the study methodology.}
\label{fig:1}
\end{figure}

This section describes the methodology adopted to conduct the study, including the research design, data collection, and analysis process. The main objective of this study is to present an overview of log issue detection research. Specifically, we aim to highlight log-related issues in software systems mentioned in the literature and the techniques used so far to identify them. For this purpose, we follow the snowballing approach proposed by Wohlin \cite{wohlin2014guidelines} to discover relevant research. Snowballing involves examining the references and citations of selected studies, helping to uncover additional research that may not be identified through standard database searches, as typically done in the traditional SLR approach outlined by Kitchenham and Charters \cite{Kitchenham}. The following steps outline our adopted approach, with Figure~\ref{fig:1} illustrating the process from defining the research questions (RQs) to discussing the results.

\subsection{Research questions}
\label{subsec4.1}

Our \textit{goal} is to advance research efforts toward log smell detection. To achieve this, we conduct an SLR, guided by clearly defined research questions that shape the focus of the review. Our RQs, along with their motivations, are presented in Table~\ref{tab:RQs}.

\begin{table}[H]
\scriptsize
\centering
\caption{Research questions addressed in this study, their identifiers (ID), and motivation.}
\resizebox{\textwidth}{!}{
\label{tab:RQs}
\renewcommand{\arraystretch}{1.3}
\begin{tabularx}{\textwidth}{|p{1cm}|p{4cm}|X|}
\hline
\multicolumn{1}{|c|}{\textbf{ID}} & \multicolumn{1}{|c|}{\textbf{Research Question}} & \multicolumn{1}{|c|}{\textbf{Motivation}} \\
\hline
RQ1 &	What types of logging smells do researchers aim to study? And how do they ground their problem?	& To understand what kinds of log smells are most commonly addressed and how each study framed and investigated its underlying research problem.\\
\hline
RQ2	& What techniques are most commonly used to detect logging smells? &	To explore which detection methods are most frequently used in the literature and assess their generalizability by examining the range of smell types they target.\\
\hline
RQ3	& How do existing studies assess the validity and effectiveness of their detection techniques for log-related issues in software systems? And what datasets are used?	& To see how researchers test their approaches and whether their findings are backed by solid evaluation and validation methods and to explore what data studies rely on, and whether certain programming languages get more attention than others, helping to see current trends and limitations in available resources.\\
\hline
RQ4	& What are the challenges, research gaps, and future directions in logging issue detection? &	To capture what problems still exist, where research is lacking, and what directions future work is likely to take.\\
\hline
\end{tabularx}
}
\end{table}

\subsection{Snowballing Process}
\label{subsec4.2}
To identify studies, we employ the snowballing technique as an alternative to traditional database searches. Snowballing involves starting with a set of seed studies that are highly relevant to the research questions. From these seed studies, we systematically review the references (backward snowballing) and the studies that cite them (forward snowballing). 
The process is iterative: each identified study is evaluated for its relevance based on the title and abstract. The selected studies are then further examined by reading the full text to assess their relevance in more detail. Additionally, the references and citations of these studies are explored to uncover more relevant research. 

Since the study addresses a specific and focused topic, the citation-based process can help identify studies that may not appear in database searches due to differences in terminology or indexing. It can also reduce the number of irrelevant papers that need to be screened, which makes the selection process more focused and manageable.

\subsection{Inclusion and Exclusion Criteria}
\label{subsec4.3}
We defined the following inclusion and exclusion criteria to determine which studies will be retained during the snowballing process:
\begin{itemize}
    \item \textbf{Inclusion Criteria:}
    \begin{itemize}
        \item[-] Primary studies presenting a model, technique, or approach for log issue detection. 
        \item[-] Studies published in English.
        \item[-] Peer-reviewed journal articles and conference papers.
        \item[-] Studies conducted in the context of software projects.
    \end{itemize}
    
    \item \textbf{Exclusion Criteria:}
    \begin{itemize}
        \item[-] Review articles, systematic reviews, and any variations or versions of the same paper (e.g., technical reports).
        \item[-] Studies that do not directly address the research questions or are not focused on log issue detection. 
 \item[-] Studies published in languages other than English.
        \item[-] Non–peer-reviewed studies to ensure that only reliable sources are included.
    \end{itemize}
\end{itemize}

\subsection{Snowballing Execution}
\label{subsec4.4}

As shown in Figure~\ref{fig:2}, the snowballing process is carried out in multiple stages:

\begin{figure}[H]
\centering
\includegraphics[width=0.5\textwidth]{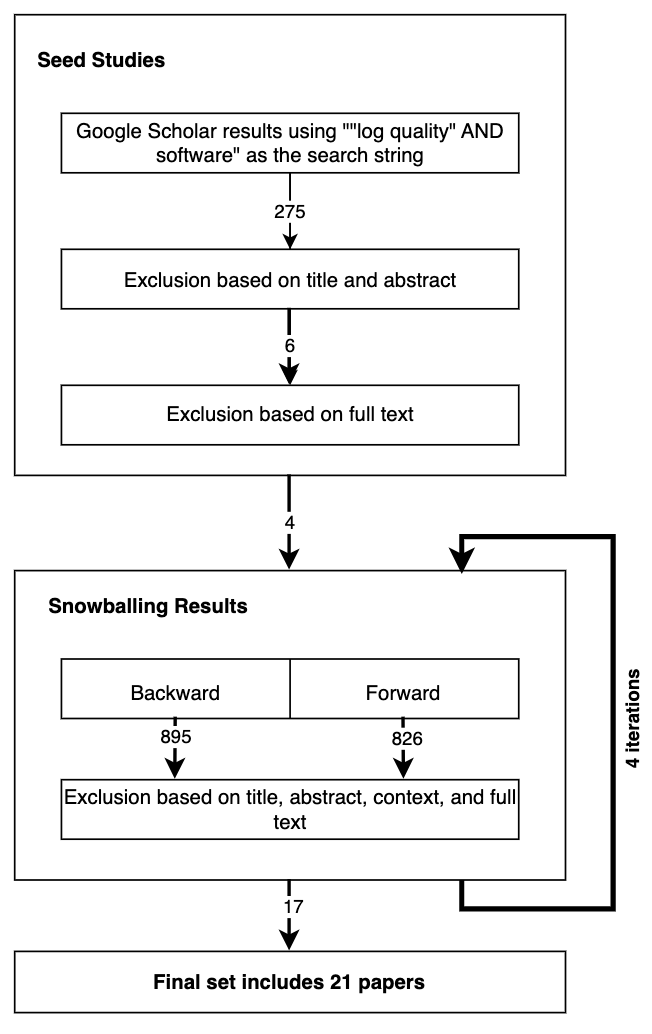}
\caption{Study selection process based on Wohlin’s snowballing approach.}
\label{fig:2}
\end{figure}

\begin{itemize}
    \item \textbf{Seed Studies:} As suggested by Wohlin \cite{wohlin2014guidelines}, we used Google Scholar to avoid bias when creating the seed studies. The search string ""logging quality" AND software" was used with AND as a boolean operator to link the two terms. The term "software" helped focus the search on studies that are specifically related to software development and maintenance. Without it, the search could have pulled in irrelevant studies from other fields where "logging" might mean something different, like tracking physical events or data in hardware. 
    
    The search term yielded 275 results; however, 6 candidate papers were found within the first three pages based on the title and abstract. The remaining pages were also browsed. After reading the full text and applying our inclusion/exclusion criteria, 4 of the 6 papers were kept as the initial seed set. Although Wohlin \cite{wohlin2014guidelines} suggests that the starting set for snowballing should not be too small, this recommendation depends on the breadth of the research area. In our case, the initial set was selected through a structured search strategy and included papers that are highly relevant to our research questions and frequently cited in the literature. Snowballing is an ongoing process, and as we track citations backward and forward, we expect to find more relevant studies. Thus, starting with a smaller but focused set remains a valid and effective approach for expanding our literature review. 
    \item \textbf{Backward Snowballing:} For each seed study, we examined its reference list to identify other relevant studies that have been cited. We first excluded any papers that did not meet our criteria or papers that have already been reviewed in previous iterations of snowballing. The remaining references were examined to determine if they were candidates for inclusion. This involved reviewing the title and reference context (i.e., text surrounding the reference in the paper) of each referenced paper. Then, the inclusion and exclusion criteria were applied to candidate papers based on their full text. This process was repeated until no new relevant studies were identified. 
    
    This process involved four iterations, during which 895 referenced papers were examined. However, only 10 papers were added to the start set, as the remaining papers were either irrelevant based on their title, context, or full text, or overlapped with those in the seed set. Consequently, the seed set comprised 14 papers in total.
    \item \textbf{Forward Snowballing:} We tracked forward citations of the seed studies to uncover more related papers. Google Scholar was used to identify the citations of each paper. The titles, abstracts, and the place citing the included paper were reviewed in the forward snowballing process. The full text was screened to make a final decision regarding the inclusion of the papers. Like backward snowballing, this process was also repeated until no new relevant studies were identified. 
    
    This process involved four iterations, during which 826 studies were examined. Based on the inclusion and exclusion criteria 7 relevant studies were added to the start set. At the end of forward snowballing, 21 relevant papers were selected as the final set for our review.
\end{itemize}

Table~\ref{tab:slr_study_details} provides a summary of the included studies. The Study ID (SID) column shows the assigned number for each paper, while the Title column presents the paper’s title. The Year column indicates when the paper was published, and the Citation Count column indicates the number of times each paper has been cited as of September 22, 2025, according to Google Scholar. The studies are organized in descending order by publication year.
\FloatBarrier

\begin{table}[htbp!]
\scriptsize
\centering
\caption{Overview of the studies included in the SLR.}
\resizebox{\textwidth}{!}{
\label{tab:slr_study_details}
\renewcommand{\arraystretch}{1.3}
\begin{tabularx}{\textwidth}{|>{\centering\arraybackslash}p{1cm}|X|>{\centering\arraybackslash}p{1cm}|>{\centering\arraybackslash}p{1.1cm}|c|}
\hline
\textbf{SID} & \multicolumn{1}{|c|}{\textbf{Title}} & \textbf{Year} & \textbf{Citation Count} & \textbf{Ref.} \\
\hline
S01 & Defects4Log: Benchmarking LLMs for Logging Code Defect Detection and Reasoning & 2025 & 0 & \cite{wang2025defects4log} \\
\hline
S02 & LogUpdater: Automated Detection and Repair of Specific Defects in Logging Statements & 2025 & 4 & \cite{zhong2025logupdater} \\
\hline
S03 & Log Sculptor: Making Logs Great Again & 2024 & 0 & \cite{kulkarni2024log} \\
\hline
S04 & When to say what: Learning to find condition-message inconsistencies & 2023 & 6 & \cite{bouzenia2023say} \\
\hline
S05 & Studying and complementing the use of identifiers in logs & 2023 & 7 & \cite{zhao2023studying} \\
\hline
S06 & On the temporal relations between logging and code & 2023 & 10 & \cite{ding2023temporal} \\
\hline
S07 & Are They All Good? Studying Practitioners' Expectations on the Readability of Log Messages & 2023 & 19 & \cite{li2023they} \\
\hline
S08 & Towards automatic detection and prioritization of pre-logging overhead: a case study of hadoop ecosystem & 2022 & 5 & \cite{zhi2022towards} \\
\hline
S09 & Qulog: Data-driven approach for log instruction quality assessment & 2022 & 7 & \cite{bogatinovski2022qulog} \\
\hline
S10 & Studying duplicate logging statements and their relationships with code clones & 2021 & 31 & \cite{li2021studying} \\
\hline
S11 & Automated evolution of feature logging statement levels using git histories and degree of interest & 2021 & 14 & \cite{tang2022automated} \\
\hline
S12 & Mobilegleak: A preliminary study on data leakage caused by poor logging practices & 2020 & 33 & \cite{zhou2020mobilogleak} \\
\hline
S13 & Three-level learning for improving cross-project logging prediction for if-blocks & 2019 & 9 & \cite{lal2019three} \\
\hline
S14 & An approach to recommendation of verbosity log levels based on logging intention & 2019 & 29 & \cite{anu2019approach} \\
\hline
S15 & An exploratory study of logging configuration practice in java & 2019 & 45 & \cite{zhi2019exploratory} \\
\hline
S16 & Studying and detecting log-related issues & 2018 & 70 & \cite{hassani2018studying} \\
\hline
S17 & An automatic approach to validating log levels in Java & 2018 & 6 & \cite{kim2018automatic} \\
\hline
S18 & Characterizing and detecting anti-patterns in the logging code & 2017 & 118 & \cite{chen2017characterizing} \\
\hline
S19 & Eclogger: cross-project catch-block logging prediction using ensemble of classifiers & 2017 & 20 & \cite{lal2017eclogger} \\
\hline
S20 & Towards just-in-time suggestions for log changes & 2017 & 113 & \cite{li2017towards}  \\
\hline
S21 & Characterizing logging practices in open-source software & 2012 & 345 & \cite{yuan2012characterizing} \\
\hline
\end{tabularx}
}
\end{table}

\subsection{Data Extraction and Synthesis}
\label{subsec4.5}
Once the 21 relevant studies were identified, we addressed each research question by extracting and analyzing related data. Table~\ref{tab:data_extraction_form} presents the data items associated with the relevant research questions. For each paper, we extracted common metadata, including paper’s titles, authors, publication year, publication type, venue, source database, and citation counts. We then extracted additional data that were designed to address all research questions.

\begin{table}[htbp!]
\scriptsize
\centering
\caption{Data extraction form used in this study.}
\resizebox{\textwidth}{!}{
\label{tab:data_extraction_form}
\renewcommand{\arraystretch}{1.3}
\begin{tabularx}{\textwidth}{
|>{\raggedright\arraybackslash}p{2cm} 
|>{\raggedright\arraybackslash}X    
|>{\centering\arraybackslash}p{2cm}|} 
\hline
\multicolumn{1}{|c|}{\textbf{Data Items}} & \multicolumn{1}{|c|}{\textbf{Description}} & \textbf{Relevant RQ} \\
\hline
Title & Title of the paper & \multirow{7}{*}{Metadata} \\
\cline{1-2}
Authors & Names of authors & \\
\cline{1-2}
Year & Year the study was published & \\
\cline{1-2}
Type & Type of publication: conference or journal & \\
\cline{1-2}
Venue & Full name of the conference or journal & \\
\cline{1-2}
Database & Source database (e.g., IEEE, ACM, Scopus) & \\
\cline{1-2}
Citation Count & Number of times the study has been cited & \\
\hline

Problem Definition & How the study grounded the issue before proposing a solution. & \multirow{2}{*}{RQ1} \\
\cline{1-2}
Smell/Issue & Specific issue or "smell" addressed &  \\
\cline{1-3}
No. of Issues & How many issues were addressed & RQ2 \\
\cline{1-3}
Challenges & Constraints/limitations reported & \multirow{2}{*}{RQ4} \\
\cline{1-2}
Future Work & Proposed future directions & \\
\hline

Technique & General method used (e.g., ML, NLP) & \multirow{3}{*}{RQ2} \\
\cline{1-2}
Model & Specific model (e.g., BERT, SVM) & \\
\cline{1-2}
Task & Task performed (e.g., classification) & \\
\hline

Code Availability & Whether the code is available & \multirow{2}{*}{Reproducibility} \\
\cline{1-2}
Dataset Availability & Whether the dataset is available & \\
\cline{1-3}
Datasets & Datasets used for training/testing & \multirow{5}{*}{RQ3} \\
\cline{1-2}
Projects Used & Software projects/repositories examined &  \\
\cline{1-2}
Programming Language & Programming languages involved & \\
\cline{1-2}
Performance Measures & e.g., precision, recall, F1, etc. & \\
\cline{1-2}
Assessment Method & Evaluation type (e.g., manual, baseline) & \\
\hline

\end{tabularx}
}
\end{table}

For RQ1, our data extraction strategy included identifying logging issues and problem definition and grounding. However, to answer the first part of this question, it was not enough to simply extract how each study describes the smell; it was also important to organize these smells in a structured way. To achieve this, we used the existing taxonomy by Saarimäki et al. \cite{saarimaki2024taxonomy} (see Section \ref{subsec2.2}) to group the smells into clear categories by mapping the smells addressed in the primary studies to their corresponding categories in the taxonomy. This approach helped us identify common patterns, differences, and gaps in the current research focus. It also made the comparison between studies easier and more meaningful.

To do this, we first reviewed Samaarki et al.’s taxonomy \cite{saarimaki2024taxonomy} , including the descriptions, examples, and related studies for each smell. Then, for each primary paper, we reviewed the reported issues and any related context (e.g., descriptions and examples). After several rounds of review and refinement, we arrived at a final mapping, which we used to answer the related research questions.

For example, Zhong et al.\cite{zhong2025logupdater} addressed four issues in their paper: statement-code inconsistency, static-dynamic inconsistency, temporal inconsistency, and readability issues (e.g., typos and capitalization). After carefully reviewing the paper, we mapped each issue to the taxonomy as follows:
\begin{enumerate}
    \item \textbf{Statement-code inconsistency} \newline $\rightarrow$ \textit{Message Madness} – wrong message
    \item \textbf{Static-dynamic inconsistency}\newline $\rightarrow$ \textit{Deceptive Variable} – wrong variable
    \item \textbf{Temporal inconsistency} \newline$\rightarrow$ \textit{Temporal Inconsistency} (direct match)
    \item \textbf{Readability issues} \newline$\rightarrow$ \textit{Message Madness} – \textit{language issues and typos}
\end{enumerate}

The second part of the question involved exploring how each study framed its research problem. This is important because there is a wide variety of logging issues, and combinations of issues, that a study could address. Therefore, we aimed to understand how researchers grounded their work and justified the specific problems they chose to investigate.

To answer RQ2, we identified the technique, model, task, and number of issues addressed by each study in order to understand how log-related problems are approached and the scope each technique is designed to handle. To capture the information needed for RQ3, we analyzed the assessment methods used in each study to identify the types of evaluation employed, including performance measures, in order to understand how consistently the effectiveness of the proposed techniques was validated and compared across the literature. For the second part of RQ3, we extracted dataset details such as the repositories or projects used and the programming languages involved, aiming to understand the basis on which researchers choose their data and how they extract and process it. Finally, to answer RQ4, we cross-analyzed the selected studies and synthesized both the challenges reported by the papers and the suggested directions for future work. We systematically identified recurring concerns across the reviewed studies and organized them accordingly. In addition to synthesizing what was explicitly stated in the papers, we also expanded on these findings by incorporating our own observations derived from the overall results of the SLR.

\section{Results}
\label{sec5}

Before answering our research questions, we first give a brief overview of the metadata of the selected papers. Table~\ref{tab:venues} presents the publication venues of the selected papers. The ratio between conference papers and journal articles was 13:8. Most of the studies were published in well-known software engineering venues, which indicates that the topic is considered relevant and important within the software engineering community. In particular, four papers were published in the \textit{International Conference on Software Engineering} (ICSE), two in \textit{IEEE’s International Conference on Software Analysis, Evolution, and Reengineering} (SANER), two in the \textit{IEEE International Conference on Software Maintenance and Evolution} (ICSME), two in the \textit{Empirical Software Engineering journal} (EMSE), and two in the \textit{IEEE/ACM International Conference on Automated Software
Engineering} (ASE). The remaining studies were each published in a distinct venue, including well-recognized venues such as \textit{ACM Transactions on Software Engineering and Methodology} (TOSEM), \textit{IEEE Transactions on Software Engineering journal} (TSE), and \textit{IEEE/ACM International Conference on Program Comprehension} (ICPC).

\begin{figure}[H]
    \centering
    \includegraphics[width=1\textwidth]{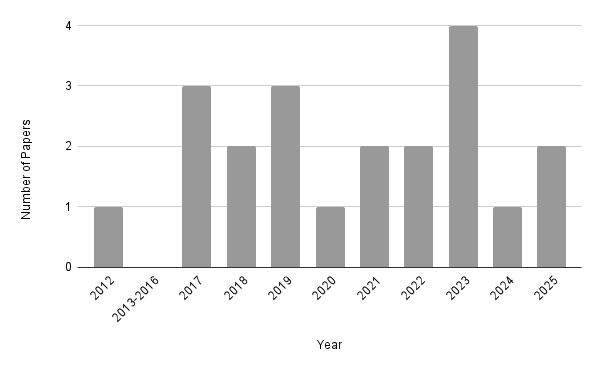}
    \caption{Yearly distribution of the 21 studies included in the review. }
    \label{fig:paper_per_year}
\end{figure}

As shown in Figure~\ref{fig:paper_per_year}, research on log smell detection has gained steady attention over the years, with a notable increase in recent years. After a slow start between 2012 and 2016, there was a rise in publications starting from 2017, with consistent contributions each year. The number of studies peaked in 2023, indicating a growing interest in the topic.

\begin{table}[H]
\scriptsize
\centering
\caption{Venues of the reviewed studies: study ID, venue name, abbreviation, type (J = journal; C = conference), and source database.}
\resizebox{\textwidth}{!}{
\label{tab:venues}
\renewcommand{\arraystretch}{2}
\begin{tabularx}{\textwidth}{|X|p{5.5cm}|X|>{\centering\arraybackslash}p{1cm}|>{\centering\arraybackslash}p{2cm}|}
\hline
\textbf{SID} & \multicolumn{1}{|c|}{\textbf{Venue}} & \textbf{Abbr.} & \multicolumn{1}{|c|}{\textbf{Type}} & \multicolumn{1}{|c|}{\textbf{Database}} \\
\hline
S02 & ACM Transactions on Software Engineering and Methodology & TOSEM & J & ACM \\
\hline
S03 & IEEE International Conference on Big Data & IEEE BigData & C & IEEE \\
\hline
S04\newline S06\newline S18 \newline S21 & International Conference on Software Engineering & ICSE & C & IEEE/ACM \\
\hline
S05 \newline S12 & IEEE International Conference on Software Analysis, Evolution and Reengineering & SANER & C & IEEE \\
\hline
S01 \newline S07 & IEEE/ACM International Conference on Automated Software Engineering & ASE & C & IEEE/ACM \\
\hline
S08 & Automated Software Engineering journal & - & J & Springer \\
\hline
S09 & IEEE/ACM International Conference on Program Comprehension & ICPC & C & IEEE/ACM \\
\hline
S10 & IEEE Transactions on Software Engineering journal & TSE & J & IEEE \\
\hline
S11 & Science of Computer Programming & SCPR & J & Elsevier \\
\hline
S13 & Journal of King Saud University-Computer and Information Sciences & JKSUC & J & Elsevier \\
\hline
S14 \newline S15  & IEEE International Conference on Software Maintenance and Evolution & ICSME & C & IEEE \\
\hline
S16 \newline S20 & Empirical Software Engineering journal & EMSE & J & Springer \\
\hline
S17 & Asia-Pacific Software Engineering Conference & APSEC & C & IEEE \\
\hline
S19 & e-Informatica Software Engineering Journal & EISEJ & J & Wroclaw University of Technology \\
\hline
\end{tabularx}%
}
\end{table}

\begin{tcolorbox}[colback=white, colframe=black, boxrule=0.5pt, sharp corners]
\textbf{Finding 1.} The trend of detecting logging smells has increased in the last few years, with a rise in publications since 2017 and a peak in 2023 and continued contributions in 2025, reflecting growing interest from the software engineering research community.
\end{tcolorbox}

Additionally, as part of our review, we examined whether the datasets and source code used in each study were publicly available to support reproducibility. Out of the 21 studies reviewed, 13 papers provided accessible datasets (e.g., S02 and S10). However, 8 papers did not include a link to their datasets (S03, S12, S13, S14, S16, S17, S19, S21). In terms of code availability, 11 studies made code publicly available with working links. 10 out of the 11 studies shared both code and data together, which helps others reproduce and build upon the work more easily (S01, S02, S04, S05, S06, S07, S08, S09, S15, S20). However, 8 papers did not provide links to code (S03, S10, S11,  S13, S14, S17, S19, S21). These results suggest that while a number of studies are taking steps toward open and reproducible research, many still do not provide the necessary resources, which limits the ability to replicate or extend their methods.

\begin{tcolorbox}[colback=white, colframe=black, boxrule=0.5pt, sharp corners]
\textbf{Finding 2.} Reproducibility is not consistently supported across studies. While 10 out of 21 studies shared both code and data, enabling replication, 8 studies did not provide datasets, and another 8 did not share their code.
\end{tcolorbox}

\subsection{RQ1: What types of logging smells do researchers aim to study? And how do they ground their problem?}
\label{subsec5.1}

To address this research question, we analyzed the 21 primary studies included in our review. Based on the extracted data, we identified two main aspects: the types of issues addressed and how each study defines its problem. The results are organized into the following sections:

\begin{table}[htbp!]
\scriptsize
\centering
\caption{Logging smell categories and facets with associated study IDs. }
\resizebox{\textwidth}{!}{%
\label{tab:smell_categories}
\renewcommand{\arraystretch}{1.3}
\begin{tabularx}{\textwidth}{|p{3.5cm}|p{5.5cm}|X|}
\hline
\multicolumn{1}{|c|}{\textbf{Log Smell/Issue}} & \multicolumn{1}{|c|}{\textbf{Facet}} & \multicolumn{1}{|c|}{\textbf{SID}} \\
\hline
\textbf{Message Madness} & missing & S03 \\
& wrong or imprecise & S01, S02, S03, S04, S07, S09 \\
& language issues/ typos & S01, S02, S03, S07, S16 \\
& duplicated message & S10 \\
\hline
\textbf{Deceptive Variable} & wrong & S01, S02 \\
& malformed output & S01, S18 \\
\hline
\textbf{Mercurial Logging Level} & incorrect & S01, S03, S09, S11, S14, S16, S17, S18 \\
& inconsistent levels & S21 \\
\hline
\textbf{Format Turmoil} & incomplete & S01, S03, S07 \\
\hline
\textbf{Landfill Logs} & too detailed, useless & S03 \\
\hline
\textbf{Undercover Identifier} & missing & S01, S05 \\
\hline
\textbf{Sleeping Guards} & missing guards & S08, S16 \\
\hline
\textbf{Logging Lost in the Wind} & missing logging (if blocks, catch blocks, exception message) & S13, S16, S19 \\
\hline
\textbf{Variable on the Edge} & nullable objects & S18 \\
& explicit cast & S18 \\
\hline
\textbf{Skeleton in the Closet} & long logging code & S18 \\
& duplicated code & S18 \\
\hline
\textbf{Other} & CO: information leak & S01, S03, S12 \\
& CO: Accidental side-effects of logging code & S01, S03 \\
& CO: Temporal inconsistency & S02, S06 \\
& CA: Logging libraries & S15 \\
& CA: Insufficient logging code maintenance & S20 \\
\hline
\end{tabularx}
}
\end{table}

\subsubsection{Issue Types}

As mentioned in Section 2, Saarimäki et al.’s log smell taxonomy \cite{saarimaki2024taxonomy} not only categorizes logging smells but also explains their possible causes and consequences. In our review, we identified 17 different smells, along with 3 types of consequences and 2 types of causes reported in the selected studies. Considering studies that address the detection of log smell causes and consequences is important because it provides useful insights for designing effective detection methods for a variety of logging issues. Understanding why a smell occurs (its cause) or what effect it may have (its consequence) can guide the development of targeted detection techniques, rather than relying only on surface-level patterns. Table~\ref{tab:smell_categories} presents the identified log smell categories, their associated facets, and the study IDs that report each smell.

\begin{figure}[H]
    \centering
    \includegraphics[width=1\textwidth]{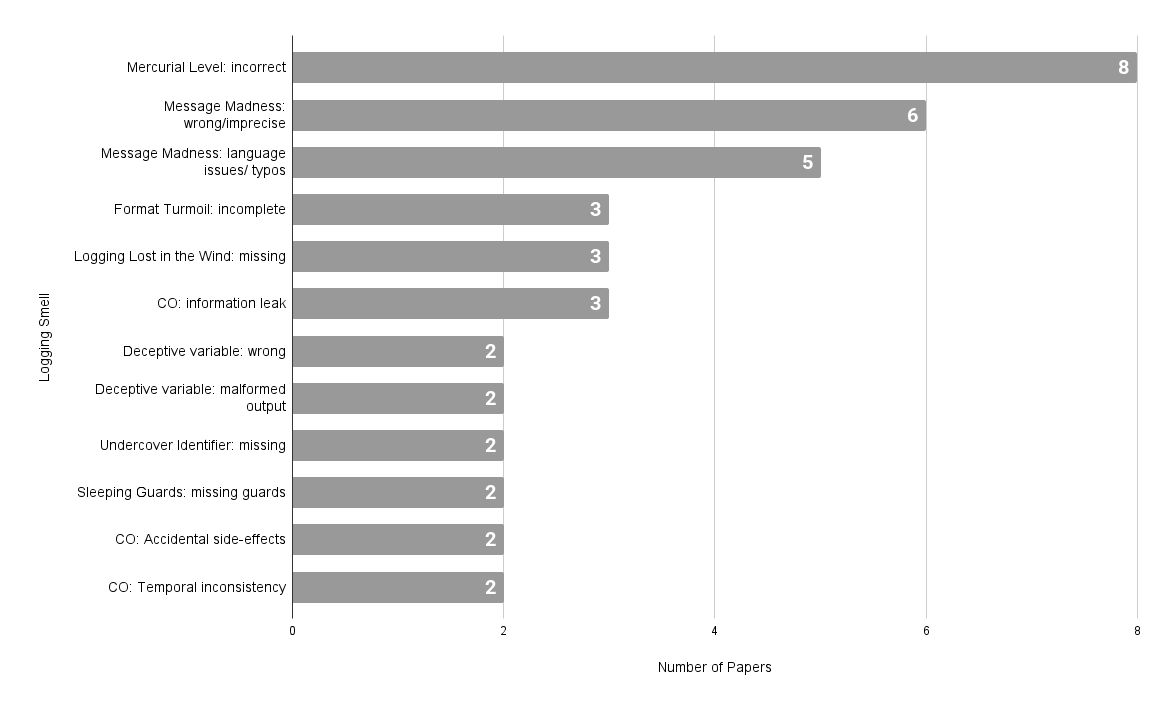}
    \caption{Distribution of logging-smells across the reviewed studies (papers per facet).}
    \label{fig:smell_vs_paper}
\end{figure}

Figure~\ref{fig:smell_vs_paper} illustrates the distribution of logging smell facets across the reviewed studies, showing the number of papers that has reported each specific smell. The most frequently reported smell facet is Mercurial Level: incorrect, appearing in 8 papers. This suggests that incorrect logging levels are a prominent and widely recognized issue in logging practices. Other commonly reported issues include Message Madness: wrong/imprecise (6 papers) and Message Madness: language issues/typos (5 papers), indicating strong attention to the quality and clarity of log messages. On the other hand, several smell facets (not shown in the chart to maintain readability), such as duplicated message, duplicated, and long logging code, appear only once, which reflects limited tool support for their detection. The studies also show some diversity in the types of smells discussed, yet only a few receive repeated attention. This uneven distribution suggests that the literature may be focused on a narrow subset of issues, leaving other potentially impactful smells less examined.

\begin{tcolorbox}[colback=white, colframe=black, boxrule=0.5pt, sharp corners]
\textbf{Finding 3.} The literature tends to concentrate on a small set of recurring log smell types, particularly incorrect log levels and issues related to log messages. While these smells are reported in multiple studies, most other smell types appear in only one or two papers. This suggests that current research efforts are not evenly distributed across the full spectrum of known log quality issues, leaving many smells insufficiently studied or overlooked in existing detection approaches.
\end{tcolorbox}

\subsubsection{Problem Definition and Grounding}
In addition to identifying the types of log issues addressed in the examined literature, we analyzed how each study frames and investigates its underlying research problem. Through this analysis, two primary approaches to problem definition emerged:

\textbf{First: Empirical or Exploratory Study} consists of studies that define their problem based on diverse investigative methods. These include preliminary analyses employing various techniques (e.g., S01, S02, S04, S09, S16), qualitative examinations of real-world logging code issues (e.g., S06, S10, S18), and large-scale quantitative characterizations of logging practices (e.g., S08, S21). Some works adopt mixed-method approaches, combining interviews, surveys, and manual inspection (e.g., S07). These empirical analyses enable researchers to uncover logging issues, their nature, and root causes, encompassing areas such as identifying defect types by studying issue reports from tracking systems (i.e., JIRA) (S16); exploring key logging aspects like identifiers (S05) and temporal relations (S06); defining properties related to log quality and readability (S09, S07); understanding motivations behind logging code changes using commit history (S06, S20, S18); examining logging guard and configuration practices (S08, S15); and demonstrating the importance of logging in software development (S14, S21).

\textbf{Second: Based on Literature Gaps} includes a minority of studies that select their target smells or issues primarily by referencing existing gaps in prior work without conducting explicit empirical investigations themselves. Only 3 out of the 21 studies (14\%) fall within this category (see Table~\ref{tab:problem_definition}). This finding emphasizes that most research in the logging domain proactively grounds its problem formulation in empirical evidence, which is particularly critical for log smell detection where understanding real-world logging behavior is essential to producing effective and relevant detection techniques.

\begin{table}[htbp!]
\scriptsize
\centering
\caption{Types of problem-definition approaches identified from the reviewed studies.}
\resizebox{\textwidth}{!}{
\label{tab:problem_definition}
\renewcommand{\arraystretch}{1.3}
\begin{tabularx}{\textwidth}{|p{7cm}|X|}
\hline
\textbf{Problem Definition and Grounding Approaches} & \multicolumn{1}{|c|}{\textbf{SID}} \\
\hline
Empirical or Exploratory Study & S01, S02, S04, S05, S06, S07, S08, S09, S10, S11, S12, S14, S15, S16, S17, S18, S20, S21 \\
\hline
Based on Literature Gaps & S03, S13, S19 \\
\hline
\end{tabularx}
}
\end{table}

\begin{tcolorbox}[colback=white, colframe=black, boxrule=0.5pt, sharp corners]
\textbf{Finding 4.} Most of the reviewed studies (85\%) define log smell problems based on empirical analysis (i.e., through various preliminary, qualitative, and quantitative investigations), highlighting the importance of analyzing actual software artifacts such as logging code, commit history, and issue reports, as this provides the necessary foundation for designing effective detection techniques.
\end{tcolorbox}

\subsection{RQ2: What techniques are most commonly used to detect logging smells?}
\label{subsec5.2}
In this section, we address the second research question by identifying and categorizing the different techniques used to detect logging issues in the selected studies. This includes examining the models employed, the specific tasks addressed (e.g., binary classification, clustering), and the number of smell types handled by a single solution, which reflects its level of generalization.

\begin{figure}[H]
\centering
\includegraphics[width=1\textwidth]{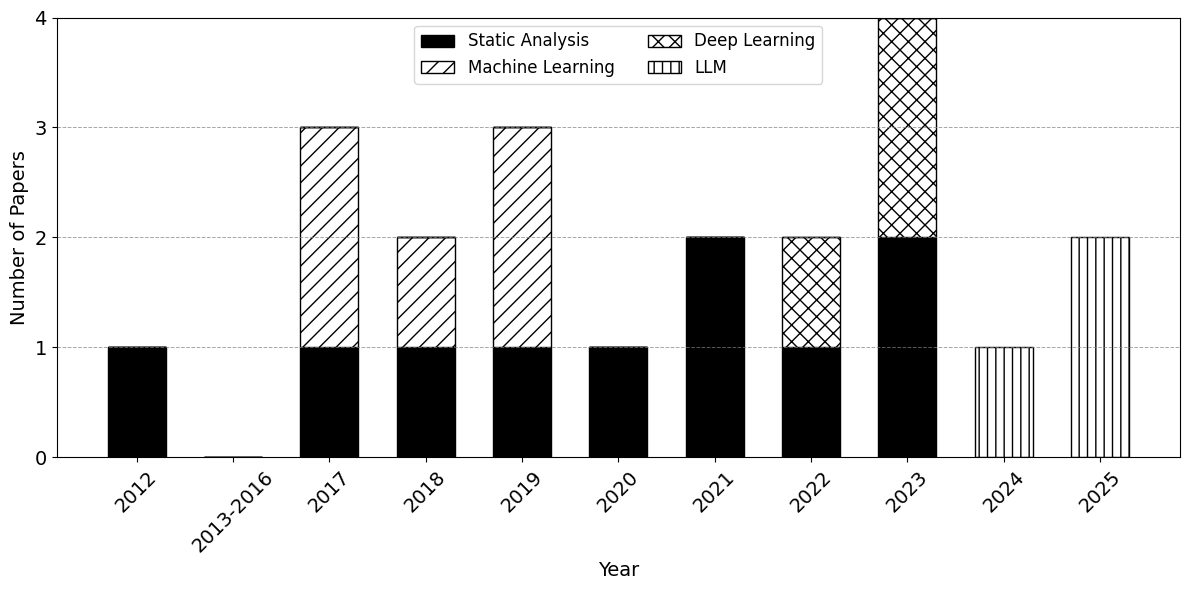}
\caption{Yearly distribution of techniques utilized in the reviewed studies.}
\label{fig:5}
\end{figure}

\begin{table}[htbp!]
\scriptsize
\centering
\caption{Detection-technique categories and their reported strengths and weaknesses.}
\resizebox{\textwidth}{!}{%
\label{tab:detection_techniques}
\renewcommand{\arraystretch}{1.3}
\begin{tabularx}{\textwidth}{|p{3cm}|X|p{4cm}|p{4cm}|}
\hline
\multicolumn{1}{|c|}{\textbf{Technique Category}} & \multicolumn{1}{|c|}{\textbf{SID}} & \multicolumn{1}{|c|} {\textbf{Reported Strengths}} & \multicolumn{1}{|c|}{\textbf{Reported Weaknesses}} \\
\hline
Static Analysis & S05\newline S06\newline S08\newline S10\newline S11\newline S12\newline S15\newline S16\newline S18\newline S21 &
  - Excels at accurately detecting well-defined structural issues (S06, S10) \newline
  - Scales to large codebases (S10, S21) \newline
  - Produces explainable and deterministic results (S16, S18) \newline
  - Automation and low cost (S11)

&
  - Poor generalization beyond predefined patterns (S10, S18)\newline
  - No runtime behavior awareness (S21)\newline
  - Dependent on quality of underlying technologies (S06, S12) \newline
  - High false positives (S06, S16)
 \\
\hline
\multirow{2}{*}{\parbox[t]{3cm}{Machine Learning}}
& S07\newline S13\newline S14\newline S17\newline S19\newline S20 &
- Integrates diverse features (syntactic, semantic, historical) (S09, S14, and S17)\newline
- Detects subtle issues missed by rule-based tools (S09, S14) \newline
- Adaptable to various architectures and tasks (S04, S19, and S20)\newline
- Can generate synthetic training data (S04, S14)

&

- Requires large, high-quality datasets (S13, S19)\newline
- Limited reasoning (S14)\newline
- Some models lack cross-project generalization (S13, S19)\newline
- Prone to bias from manual labels (S07, S17)\newline
 \\
\cline{1-2}
Deep Learning & S04\newline S07\newline S09 & & \\
\hline
Large Language Models & S01\newline S02\newline S03 &

- Can detect a broad range of logging defects (S01, S02, and S03)\newline
- Prompts can be enriched with defect-specific or project-specific knowledge and logging guidelines (S01, S03) \newline
- Can reason over natural language in log statements (S01, S02)\newline
 - Can be quickly adapted to new projects or defect types without retraining (S01, S03)
&
- Running on large codebases can be slow and computationally expensive (S02, S03) \newline
- Correct predictions may be accompanied by incomplete or incorrect explanations (S01) \newline
- Prone to misclassifying correct log statements as defective without additional filtering or checking (S01, S02) \newline
- Often lack structural code awareness, limiting interpretation of control/data flow (S01, S02) \\
\hline
\end{tabularx}
}
\end{table}

\subsubsection{Detection Techniques}

Four main techniques are used in the reviewed studies to detect logging issues: static analysis, machine learning, deep learning, and LLMs. Figure~\ref{fig:5} illustrates their yearly distribution, while Table~\ref{tab:detection_techniques} summarizes these techniques with corresponding studies and their reported strengths and weaknesses. Below, we provide an overview of each technique and how it is used in the reviewed papers.

 Static analysis-based studies aim to examine the source code without executing it, in order to detect issues related to logging practices. In these studies, researchers utilize several techniques, such as abstract syntax tree (AST) analysis, to identify structural issues including patterns and anti-patterns in logging code (S10, S16, S18). Some approaches also rely on control and data flow analysis, pointer analysis, and inter-procedural techniques to capture the logical behavior and dependencies in the code (S05, S08). A few tools apply taint flow analysis to check whether sensitive data is unintentionally exposed through log statements (S12). Moreover, some static analysis-based studies incorporate linguistic processing techniques such as part-of-speech (POS) tagging and dependency parsing (S06), as well as heuristic-based techniques and static rules (S11).

Another category of techniques comprises machine learning-based approaches, which aim to detect and classify logging issues by learning patterns from labeled or unlabeled data. These studies often employ traditional supervised learning algorithms such as Random Forest, Naïve Bayes, Logistic Regression, and Decision Trees to predict log quality attributes like verbosity levels or the presence of logging anomalies (S07, S13, S17, S19, S20). For example, S19 leverages a binary classification task and applies ensemble methods, including bagging and majority voting, to effectively combine multiple classifiers and enhance prediction performance. In addition, the approach in S14 applies unsupervised learning techniques, specifically clustering methods, to group similar logging instances based on their contextual features and to detect inconsistencies in their verbosity levels.
% enhanced by MinHash and locality-sensitive hashing (LSH),

Moreover, deep learning-based studies represent techniques that utilize neural network architectures to capture complex patterns in logging data. Some studies employ Bidirectional Long Short-Term Memory (BiLSTM) networks for binary classification of log messages into  \textit{adequate} or \textit{inadequate} in terms of readability (S07). CMI-Finder (S04) employs three neural models for detecting condition--message inconsistencies: (i) a binary classification model that encodes concatenated condition--message pairs using a BiLSTM and predicts the probability of inconsistency; (ii) a triplet-loss model that embeds conditions and log messages separately into a shared vector space to distinguish consistent from inconsistent pairs based on distance; and (iii) a fine-tuned CodeT5~\cite{wang-etal-2021-codet5} transformer that leverages pre-trained code understanding to classify pairs as \textit{consistent} or \textit{inconsistent}.
 
In S09, QuLog uses a Transformer encoder with multi-head self-attention to model token dependencies in log messages. It addresses two tasks: (i) log-level prediction, a multi-class classification with three classes—\textit{info}, \textit{warning}, \textit{error}; and (ii) linguistic-structure assessment, a binary classification with two classes—\textit{sufficient} and \textit{insufficient} linguistic richness. Explainable-AI techniques highlight influential tokens to help developers improve log clarity and correctness.

Feature engineering in machine learning and deep learning is sometimes supported by natural language processing (NLP) techniques, such as stemming, stopword removal, Term Frequency – Inverse Document Frequency (TF-IDF) conversion, and Word2Vec embeddings, to convert log messages into informative representations for training (e.g., S09, S13, S17).

Recently, LLMs are employed to enhance the analysis and generation of log statements through their advanced understanding of natural language and code. For example, in S01, several LLMs (i.e., GPT-4o \cite{openai2023gpt4}, Llama3.3 \cite{llama3}, DeepSeek-R1 \cite{deepseekr1}, and Qwen2.5-72B \cite{qwen2025}) are leveraged to detect and reason about logging code defects through a benchmark framework called Defects4Log. The authors build a comprehensive taxonomy of seven defect patterns, such as \textit{readability issues}, \textit{variable issues}, \textit{logging level misuse}, \textit{semantic inconsistencies}, \textit{sensitive information leaks}, \textit{insufficient information}, and \textit{performance problems}, and compile a dataset of 164 developer-verified real-world logging defects. They evaluate various prompting strategies (e.g., Chain of Thought (CoT) \cite{wei2022cot}) and contextual inputs (e.g., control flow) to test LLMs’ ability to detect and explain these defects. %, and show that incorporating prior knowledge about the defect scenarios improves detection accuracy. 

Additionally, LogUpdater (S02) is a two-stage framework that detects and repairs four types of logging statement defects: \textit{statement-code inconsistency}, \textit{static-dynamic inconsistency}, \textit{temporal inconsistency}, and \textit{readability issues}. In the detection phase, it fine-tunes a similarity-based classifier and evaluates its performance against several LLM baselines, including Claude3.5-Sonnet \cite{anthropic2024claude35}, GPT-3.5 \cite{openai2022gpt35}, GPT-4o \cite{openai2023gpt4}, DeepSeek-Coder-v2 \cite{zhu2024deepseekcoder}, and DeepSeek-R1 \cite{deepseekr1}, using few-shot prompting to detect defects.

%In the detection phase, it fine-tunes a similarity-based classifier on 32K synthetic defective log samples created via mutation strategies, including LLM-based semantic mutations using GPT-4o \cite{openai2023gpt4}. For comparison, the study also evaluates several LLM baselines, including Claude3.5-Sonnet \cite{anthropic2024claude35}, G

Log Sculptor (S03) also uses LLMs like GPT-3.5 \cite{openai2022gpt35} and GPT-4 \cite{openai2023gpt4} with a prompt-based approach to identify and improve poor log statements in code. It first detects log statements using a deterministic pre-processing step, then analyzes them based on eight common poor logging practices and classifies them into four classes (\textit{Required}, \textit{Should be changed}, \textit{Should be removed}, and \textit{Absent}), using best practices and project-specific guidelines as additional context in an instruction prompt.

\begin{tcolorbox}[colback=white, colframe=black, boxrule=0.5pt, sharp corners]
\textbf{Finding 5.} From 2012 to 2021, static analysis and traditional machine learning dominated the field. However, starting in 2022, there has been a gradual increase in the adoption of deep learning techniques. Most notably, 2024 marks the introduction of Large Language Models, signaling the latest advancement in logging smell detection.
\end{tcolorbox}

\subsubsection{Multi-Smell Detection}
Although multiple studies explore more than one smell (i.e., S01, S02, S03, S07, S09, S16, S18) (see Table~\ref{tab:smell_categories}), not all of them use a single detection solution to identify multiple smells. For example, while S16 addresses four smells, it develops a dedicated checker for each one. Similarly, S18 investigates six smells but implements separate modules for detecting each. However, both studies develop a single tool that integrates these individual detection components. Table~\ref{tab:multi_smell_labels} shows the studies that address multiple logging smells and the labels used to classify each one. While both S07 and S09 target multiple smells, their approaches are based on separately classifying each type of smell. For instance, S07 classifies each smell type as either "\textit{adequate}" or "\textit{inadequate}." Similarly, S09 classifies logging levels independently from the assessment of logging messages. The advantage of this approach is that each logging statement can be assessed based on a variety of smells, allowing for a more comprehensive evaluation of its quality. However, this also means that separate models or components must be maintained for each smell type, which can increase the complexity of implementation and reduce scalability across larger or more diverse codebases.

On the other hand, S01, S02, and S03 follow approaches capable of detecting multiple smells using a single tool or framework employing multi-class classification via LLM techniques. However, S01 and S02 do not account for situations where a log statement may exhibit more than one smell (i.e., belong to multiple labels), which may limit its effectiveness in real-world scenarios where logging issues often co-occur. In contrast, the framework in S03 could be used to identify multiple issues simultaneously within a single logging statement, demonstrating a more comprehensive detection capability.

\begin{table}[htbp!]
\scriptsize
\centering
\caption{Studies that address multiple smells and their used labels.}
\resizebox{\textwidth}{!}{
\label{tab:multi_smell_labels}
\renewcommand{\arraystretch}{1.3}
\begin{tabularx}{\textwidth}{|>{\centering\arraybackslash}p{3cm}|p{6cm}|X|}
\hline
\textbf{SID} & \multicolumn{1}{|c|}{\textbf{Smell}} & \multicolumn{1}{|c|}{\textbf{Class Labels}} \\
\hline

\multirow{8}{*}{\textbf{S01}} 
& Message Madness – wrong or imprecise  & \multirow{2}{*} {RD: Readability Issues} \\
\cline{2-2}
& Message Madness – language issues like typos & \\
\cline{2-3}
& Deceptive variable – wrong   & \multirow{2}{*} {VR: Variable Issues}  \\
\cline{2-2}
& Deceptive variable – malformed output & \\
\cline{2-3}
& Mercurial logging level – Incorrect & LV: Logging Level Issues  \\
\cline{2-3}
& Deceptive variable – wrong   & \multirow{2}{*} {SM: Semantics Inconsistent}  \\
\cline{2-2}
& Message Madness – wrong or imprecise & \\
\cline{2-3}
& CO: information leak & SS: Sensitive Information  \\
\cline{2-3}
& Format turmoil – incomplete   & \multirow{2}{*} {IS: Insufficient Information}  \\
\cline{2-2}
& Undercover Identifier - missing & \\
\cline{2-3}
& CO: Accidental side-effects of logging code & PF: Performance Issues   \\
\cline{2-3}
& No Smell & Non-defective \\
\hline

\multirow{5}{*}{\textbf{S02}} 
& Message Madness – wrong & Statement-code inconsistency \\
\cline{2-3}
& Deceptive variable – wrong & Static-dynamic inconsistency \\
\cline{2-3}
& CO: temporal inconsistency & Temporal inconsistency \\
\cline{2-3}
& Message Madness – language issues like typos & Readability inconsistency \\
\cline{2-3}
& No Smell & Non-defect \\
\hline
\multirow{9}{*}{\textbf{S03}} 
& Message madness – missing & \multirow{2}{*}{Absent}  \\
\cline{2-2}
& Format turmoil – incomplete & \\
\cline{2-3}
& Mercurial logging level – Incorrect & \multirow{4}{*}{Should be changed} \\
\cline{2-2}
& CO: information leak & \\
\cline{2-2}
& Message madness – imprecise & \\
\cline{2-2}
& Landfill logs – Too detailed & \\
\cline{2-3}
& Landfill logs – Useless & \multirow{2}{*}{Should be removed} \\
\cline{2-2}
& CO: Accidental side-effects of logging code & \\
\cline{2-3}
& No Smell & Required \\
\hline
\multirow{3}{*}{\textbf{S07}} 
& Format turmoil – incomplete & Adequate or inadequate \\
\cline{2-3}
& Message madness – imprecise & Adequate or inadequate \\
\cline{2-3}
& Message Madness – language issues & Adequate or inadequate \\
\hline
\multirow{2}{*}{\textbf{S09}} 
& Mercurial logging level – Incorrect & Info, warning, or error \\
\cline{2-3}
& Message madness – imprecise & Sufficient or insufficient \\
\hline
\end{tabularx}
}
\end{table}

\begin{tcolorbox}[colback=white, colframe=black, boxrule=0.5pt, sharp corners]
\textbf{Finding 6.} Approaches to detecting multiple logging smells vary across studies. Some use separate models or components for each smell type, while others employ unified solutions capable of handling multiple smells within a single solution. Recent studies show that LLMs can act as such unified solutions.
\end{tcolorbox}

\subsection{RQ3: How do existing studies assess the  validity and effectiveness of their detection techniques for log-related issues in software systems? And what datasets are used?}
\label{subsec5.3}
To answer this research question, we analyzed the selected studies based on the methods they use to evaluate and/or validate their results. In addition, we synthesized the projects used as datasets and identified the programming languages in which they were implemented.

\subsubsection{Assessment Methods}
Table~\ref{tab:assessment-methods} shows the types of assessment methods, their description for clarification, and studies that apply them. The analysis of assessment methods revealed several recurring trends in how detection techniques are evaluated across different stages of their development and application in the literature. A large proportion of studies (17 out of 21) employ manual validation by authors, which reflects a practical but subjective approach wherein researchers assess the correctness of their outputs based on expert judgment. 

\begin{table}[htbp!]
\scriptsize
\centering
\caption{Types of assessment methods used in the reviewed studies.}
\resizebox{\textwidth}{!}{
\label{tab:assessment-methods}
\renewcommand{\arraystretch}{1.3}
\begin{tabularx}{\textwidth}{|p{3cm}|p{5cm}|X|}
\hline
\multicolumn{1}{|c|}{\textbf{Assessment Type}} & \multicolumn{1}{|c|}{\textbf{Description}} & \multicolumn{1}{|c|}{\textbf{SID}} \\
\hline
Manual Validation by Authors & Researchers manually review outputs to judge correctness. & S02, S03, S04, S05, S06, S07, S08, S09, S10, S11, S12, S15, S16, S18, S19, S20, S21 \\
\hline
Comparison with a Baseline & Results are compared to existing tools or techniques & S01, S02, S04, S08, S09, S10, S13, S14, S16 \\
\hline
Evaluation Against Ground Truth & Results are compared to annotated or known correct data. & S01, S02, S03, S06, S07, S08, S09, S10, S13, S16, S18, S19, S20 \\
\hline
Testing on Real-World Projects & Techniques are applied to real software projects for practical testing. & S01, S02, S03, S04, S05, S06, S07, S08, S09, S10, S11, S12, S13, S14, S15, S16, S17, S18, S19, S20, S21 \\
\hline
Performance Metrics & Standard metrics (e.g., precision, recall, F1, AUC, accuracy) are used to measure performance. & S01, S02, S03, S04, S06, S07, S08, S09, S10, S11, S13, S14, S16, S17, S18, S19, S20 \\
\hline
Developer Feedback & Results are reported to actual developers (e.g., via Pull Requests or issue trackers) and confirmation is tracked. & S04, S06, S08, S09, S10, S11, S14, S15, S16, S18, S21 \\
\hline
\end{tabularx}
}
\end{table}

Testing on real-world projects is also common across the 21 studies, however, the selected projects vary in size, domain, and popularity, which may affect the generalizability and comparability of results (See section \ref{subsec5.3.2}). Furthermore, quantitative performance metrics such as precision, recall, F1-score, accuracy, and Area Under the ROC Curve (AUC) are used in 17 studies, especially those involving machine or deep learning models. However, the choice of metrics differed across studies, indicating a lack of consistency in how detection performance is evaluated. Some report precision, recall, accuracy, and F-measure to evaluate the effectiveness of their detection models (e.g., S02, S07, S09), while others emphasize the use of AUC to assess the discriminative ability of classifiers in distinguishing between positive and negative cases (e.g., S04, S13, S20).

Among the studies, evaluation against ground truth, such as annotated datasets or known issue fixes, is used in 13 cases. This method offers objective assessment, although it often depends on the quality and completeness of the underlying data. Also, the types of ground truth vary widely depending on the research questions and the aspect of logging being studied, such as real-world developer changes or fixes (e.g., S02), manually created or labeled datasets derived from manual analysis and expert evaluation of developer written logging statements (e.g., S10), and synthetic data generated through techniques like mutation or LLMs (e.g., S02, S04). 

Developer feedback appeared in 11 studies, usually through issue reports or pull requests, showing that some researchers made the extra effort to confirm their results with actual practitioners. However, only 8 studies conduct comparative evaluations with existing baselines, and even those baselines vary. For example,  while S09 and S16 both tackle the problem of inappropriate log levels, S09 evaluates its approach against baselines like DeepLV \cite{li2021deeplv} and Support Vector Machine, while S16 evaluates its specific log level checker against a prior statistical model \cite{li2017log}. 
Overall, while the practice of using multiple assessment methods is common, the specific approaches, strategies, and metrics differ across studies.

\begin{tcolorbox}[colback=white, colframe=black, boxrule=0.5pt, sharp corners]
\textbf{Finding 7.} The use of multiple assessment methods is widespread, including manual validation by authors, comparison with a baseline, evaluation against ground truth, testing on real-world projects, performance metrics, and developer feedback. However, the use of different performance metrics makes it difficult to consistently compare results across studies.
\end{tcolorbox}

\subsubsection{Datasets} \label{subsec5.3.2}
The analysis of datasets used across the selected studies revealed that, although many rely on publicly available and open-source real-world projects such as Apache Tomcat \cite{apacheTomcat}, Hadoop \cite{apacheHadoop}, and CloudStack \cite{apacheCloudStack}, these datasets are not immediately suitable for log issue detection. Instead, researchers typically follow a series of steps, including project and repository selection (based on specific criteria, when stated), collection of logging statements or their changes, and manual labeling. Moreover, the methods used to perform these steps vary across studies, which often results in differing numbers of projects and log statements used.

In terms of project selection strategies, studies often apply specific criteria when choosing projects for dataset construction, depending on their goals. Project popularity and maturity are among the most frequently used criteria, as observed in studies such as S01, S02, S08, S09, S14, S15, S17, S18, and S21. These works typically consider factors like the number of GitHub stars, development history, and sustained maintenance activity, often measured by project age and commit frequency, to ensure that the selected systems are both reliable and representative. Additionally, some studies intentionally select projects from diverse domains (e.g., S05, S14, S15, S20), while others, such as S17, focus on a single domain to maintain consistency. A few studies, including S10 and S06, base their selection on prior usage in log-related research, whereas others reuse or build upon datasets from earlier work (e.g., S03, S05, S07, S16). In contrast, some studies like S04 and S13 do not clearly report their project selection criteria, which may limit reproducibility. Overall, although the selection strategies vary, popularity, maturity, and domain relevance appear to be the most widely adopted characteristics in project selection.

Next, many studies collect logging statements directly from the source code of selected projects, often using static analysis and/or regular expressions (e.g., S05, S06, S10, S13, and S19). Studies that analyze log changes or issues collect data from version control repositories (e.g., Git or SVN) and issue tracking systems (e.g., JIRA) (e.g., S06, S15, S16, S18, S20) via commit histories and issue reports. For example, S16 uses a keyword-based heuristic ("log", "logging", "logger") to filter log-related issues from JIRA reports for further study. 

Similarly, preprocessing steps vary widely based on the type of data and the study's objective. For instance, textual data, such as log messages or code snippets, often undergo preprocessing like filtering (e.g., S18), tokenization (e.g., S07), POS tagging (e.g., S09), or conversion to numerical representations like TF-IDF (e.g., S13 and S19). Another step that varies across studies is dataset labeling, which is typically performed by human experts, often the researchers themselves. Most importantly, the labels are usually custom-defined to reflect the specific objectives of each study, rather than based on any shared or standardized scheme (e.g., \textit{adequate/inadequate}, \textit{sufficient/insufficient} linguistic structure) (see Table~\ref{tab:multi_smell_labels}). 

\begin{tcolorbox}[colback=white, colframe=black, boxrule=0.5pt, sharp corners]
\textbf{Finding 8.} Many studies use open-source projects, but they require significant customization for log issue detection. Project selection, data extraction, and labeling practices vary, often tailored to specific research goals. This lack of standardization highlights the need for shared benchmarks and reusable datasets in the field.
\end{tcolorbox}

\begin{figure}[H]
    \centering
    \includegraphics[width=1\textwidth]{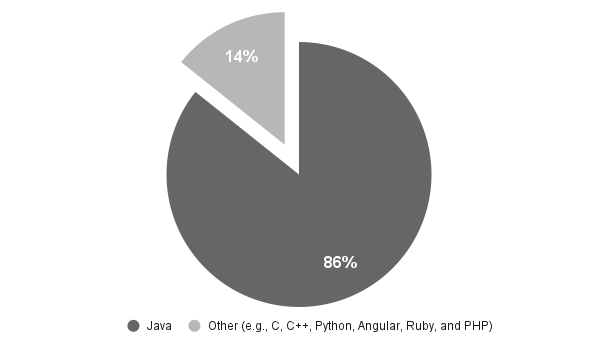}
    \caption{Distribution of programming languages across studied projects.}
    \label{fig:PL}
\end{figure}

In terms of programming languages, we found that approximately 86\% of the reviewed studies focus on Java-based projects (e.g., Hadoop \cite{apacheHadoop}, Zookeeper \cite{apacheZooKeeper}, and HBase \cite{apacheHBase}), particularly those maintained by the Apache Software Foundation (ASF) \cite{apacheHomepage}. In contrast, only a few studies address logging issues in languages other than Java. Specifically, S04 reports one smell in Python, and S09 reports two smells in Python, Angular, Ruby, and PHP. Additionally, S21 reports one smell in C/C++. This strong concentration on Java indicates a clear language bias, which may limit the generalizability and applicability of existing detection techniques across different programming environments (see Figure~\ref{fig:PL}).

To address this limitation, there is a need to broaden the scope of future research. In addition to widely studied open-source Java systems, researchers should consider other parts of the Java ecosystem, such as Android-based projects. Moreover, further attention should be given to closed-source systems and projects written in other programming languages, as logging practices are likely to differ across languages, platforms, and application domains.

\begin{tcolorbox}[colback=white, colframe=black, boxrule=0.5pt, sharp corners]
\textbf{Finding 9.} The majority of studies (approximately 86\%) focus on Java-based projects, revealing a strong language bias that needs to be addressed.
\end{tcolorbox}

The results of RQ3 may suggest that the field lacks established benchmarks, which  makes it difficult to compare results. This is evident in several observations: 
\begin{enumerate}
    \item Several previous studies treat the logging statements written by developers as the ground truth for evaluation.
    \item Studies use different log statements, commits, and issue reports from different projects, and these are not standardized.
    \item Limited comparisons with existing baselines, and those used are inconsistent.
    \item Studies use different or inconsistent evaluation metrics; some use precision/recall, others use AUC.
    \item While some studies aim to detect incorrect log levels and others focus on message inconsistency or duplication, there is no unified task definition that standardizes how such problems are framed or assessed. This makes it challenging to compare methods across studies in a consistent way.
\end{enumerate}

\begin{tcolorbox}[colback=white, colframe=black, boxrule=0.5pt, sharp corners]
\textbf{Finding 10.} The lack of shared datasets, consistent baselines, and reusable evaluation protocols suggests that the field does not yet have well-established benchmarks for detecting log-related issues. This situation makes it harder to reproduce results, compare findings fairly, and maintain consistency across studies. Therefore, there is a clear need for community-agreed benchmarks and publicly available datasets to support more reliable and systematic research in this area.
\end{tcolorbox}

\subsection{RQ4: What are the challenges, research gaps, and future directions in logging issue detection?
}
\label{subsec5.4}
In this section, we discuss the challenges and future research opportunities identified through our review of log issue detection studies.
\subsubsection{Challenges}
This section outlines the main challenges faced in log issue detection studies, revealing research gaps that create opportunities for improvement.
\begin{itemize}
  \item \textbf{Lack of Standardized Guidelines} 
  
  One major challenge in logging is the lack of clear, standardized guidelines. There are no universally accepted rules for what makes a "good" or "bad" log statement, leading to inconsistencies across projects (S14, S18, S07). As highlighted in S14, even major companies like Microsoft do not provide project-specific instructions for logging behavior. This lack of standardization makes it harder for developers to ensure consistency and complicates the automation of logging quality assessment.

Although, several studies discuss logging practices that can guide practitioners when writing log statements. For example, \cite{gu2022logging} and \cite{batoun2024literature} highlight the importance of avoiding excessive or sparse logging by balancing the amount of information with the cost of storing and processing it. They also mention choosing suitable log levels and including enough context in messages, such as variable values, identifiers, and error details. In addition, they suggest reviewing and updating old or redundant log statements and placing logs in meaningful parts of the code. Following such practices during development can help reduce common logging smells and improve the long-term quality of logs.

%Leveraging such studies can help move toward more consistent research. A first step is to develop and promote community-accepted logging standards and checklists to guide practitioners and reduce inconsistencies in practice. Building on that, another step is to create datasets with labeled examples of problematic and well-formed log statements, which can support empirical evaluation  and allow researchers to assess and refine their techniques on shared benchmarks. Finally, integrating automated tooling that checks compliance with these guidelines can support their use in real projects.

\item \textbf{Diversity of Logging Defects}

Logging issues come in many forms, including incorrect verbosity levels (S16), duplicate messages (S10), readability issues (S07), inconsistent logging practices (S06, S04, S14), and missing logging statements (S16). Since logging problems are so diverse, they require multiple analysis techniques to detect and fix effectively, ranging from static code analysis and data flow analysis to natural language processing for semantic understanding. %Therefore, it is still an open challenge to create a one-size-fits-all solution, which could be addressed by leveraging the advanced capabilities of LLMs.

\item \textbf{Log Statement Extraction} 

In the beginning stages, it can be a challenge to detect log statements  from the source code for an experiment. For example, in S13 authors created 26 regular expressions to find all the log statements. Manual inspection revealed that all the types of log statements were identified, but there was still a possibility that some types of log statements were missed. Also, using GPT-4, Log Sculptor (S03) initially detected only 10.6\% of the log statements but a deterministic pre-processing step improved the detection rate to 90.6\%. This underscores the importance of pre-processing, even when utilizing advanced techniques such as LLMs. Therefore, developing robust parsers combined with advanced techniques is needed to help detect and extract log statements, especially in cloned or complex code.

\item \textbf{False Positives and False Negatives} 

Many methods suffer from high false positive rates (detecting non-issues as issues) or false negatives (missing actual problems). Tools relying on static code analysis, like LCAnalyzer (S18) for detecting logging anti-patterns and TempoLo (S06) for temporal inconsistencies, can generate false positives because they analyze code without executing it. For example, TempoLo may not detect the correct main verb, an error caused by a natural language processing tool (i.e., spaCy \cite{honnibal2017spacy}). In addition, some studies like S10 might employ strict rules, such as removing highly frequent words upon detecting log messages, which can inadvertently lead to false negatives. Also, QuLog (S09) highlights that even with deep learning methods, misclassifications in log levels can occur, underscoring the ongoing challenge of achieving high accuracy without introducing excessive false positives or negatives. Notably, the study observed higher misclassification rates attributed to a significant overlap in n-grams. 

LogUpdater (S02), on the other hand, addresses the issue of false positives resulting from imbalanced data by using a checker LLM agent in their study, which shows improved results. Also, Log Sculptor (S03) integrates developer feedback loops for continuous improvement. Evolving such solutions to combine automated detection with various feedback sources (e.g., from developers or LLMs) and complementary techniques could help improve accuracy levels.

\item \textbf{Programming Language Constraints}

Many studies focus on Java, limiting applicability to other languages. For example, applying LogUpdater (S02) to different programming languages may impact its performance due to variations in language-specific features. Additionally, some studies rely on limited datasets, such as only three open-source Java projects (e.g., S13), making it difficult to generalize their findings. Furthermore, the majority of these studies are conducted on open-source projects, which raises concerns about the applicability of results to closed-source projects, where logging practices and constraints may differ. 

%Future work should explore multi-language log detection by developing language-agnostic approaches (e.g., using cross-language ASTs or NLP-based solutions) to understand and improve their applicability across different programming languages.

\item \textbf{Cross-Project Challenges}

Logging models trained on one project often struggle to generalize to another due to differences in vocabulary (S19), numerical feature distributions (S19), and coding conventions (S10). For example, Tomcat \cite{apacheTomcat} and Hadoop \cite{apacheHadoop} have a different number of unique exception types, with many types present in one project but not the other. To address these issues, ensemble-based models such as ECLogger (S19) have been proposed, utilizing multiple classifiers to enhance prediction accuracy across different projects. Additionally, frameworks like the three-level model CLIF (S13) aim to further improve cross-project logging prediction, highlighting ongoing efforts and the potential for advancement in this research area. 

%Building on these efforts, future work could focus on developing more adaptable models that better handle project-specific variation and improve cross-project log smell detection performance by using techniques such as feature selection, feature and instance weighting, and incorporating distance-weighted transfer learning \cite{zou2021correlation} \cite{lei2022wcm}.

\item \textbf{Data Imbalance} 

Certain log levels or logging patterns are underrepresented in training data, making them harder to predict. For example, S02 discusses that while log-related updates often include defective logging statements, these constitute only a small percentage of all logging statements. Additionally, S14 highlights data imbalance due to the scarcity of certain logged instances in comparison to others. However, S13 provides the most explicit and detailed discussion of the class imbalance problem in logging prediction, particularly emphasizing the low percentage of logged if-blocks. The imbalanced data distribution underscores the need for improved dataset balancing techniques and for optimizing datasets for log issue detection to better capture rare but critical defects.
\end{itemize}

\begin{tcolorbox}[colback=white, colframe=black, boxrule=0.5pt, sharp corners]
\textbf{Finding 11.} Challenges observed across log issues detection research are related to lack of standardized guidelines, diversity of logging defects, log statement extraction, false positives and false negatives, programming language constraints, cross-project challenges, and data imbalance.
\end{tcolorbox}

\subsubsection{Recommendations for Future Research}
Based on the study findings, the challenges presented in the previous section, and on the future directions suggested by the reviewed studies, we present several observations that can help to direct future research efforts in automated log issue detection.

\begin{itemize}
  \item \textbf{Enhancing Log Analysis Techniques}

Employing NLP could enhance log analysis by analyzing the static text of log messages for inconsistencies and readability issues. For instance, future work could build upon the techniques or temporal relationships used in TempoLo (S06) and LCAnalyzer (S18) to detect inconsistencies by incorporating more advanced NLP techniques.

According to S13 and S19 machine learning techniques are also suggested for improving cross-project logging prediction for various code constructs beyond catch-blocks, such as if-blocks, potentially by incorporating more features or levels of learning. Building on these efforts, future work could also focus on developing more adaptable models that better handle project-specific variation and improve cross-project log smell detection performance by using techniques such as feature selection, feature and instance weighting, and incorporating distance-weighted transfer learning \cite{zou2021correlation} \cite{lei2022wcm}.

In addition, LLMs show potential for enhancing log interpretability, as demonstrated in the context of Log Sculptor (S03). Future work could further explore their integration into logging tools to enable more sophisticated analysis. Also, since developing a single solution addressing a wide range of defects remains an open challenge, this limitation might be mitigated by leveraging the advanced capabilities of LLMs.

To enhance the accuracy and relevance of log analysis, studies like S10, S14, and S06 recommend that future research should explore a broader range of code contexts. Since most current analyses are static, incorporating dynamic information, such as runtime behaviors, could provide deeper insights. Additionally, analyzing the sequence of generated logs can reveal how earlier entries influence or relate to subsequent ones, offering valuable context.

 \item \textbf{Expanding Support for Multiple Programming Languages}
 
A significant future direction pointed out by S19 is to evaluate and extend the applicability of logging analysis tools to more software projects, including closed-source applications and projects written in programming languages beyond Java, such as Python, C, and C++. This would involve addressing the challenges posed by different language syntax, logging libraries, and development practices.

Future work could also consider exploring multi-language log detection by developing language-agnostic approaches (e.g., using cross-language ASTs or NLP-based solutions) to understand and improve their applicability across different programming languages.

 \item \textbf{Improving Developer Tools}
 
S15 mentioned that existing logging libraries like Log4J \cite{apacheLog4jLayouts} and Logback \cite{logbackArchitecture} offer flexibility when building logging components. However, developers have multiple needs, potentially leading to the use of more than one logging library within a single project. This can increase the effort required to produce systematic and uniform logs due to potential incompatibility or limited configurability between different libraries. Even if similar configurations are possible, maintaining consistency across multiple libraries can be challenging \cite{saarimaki2024taxonomy}. This reinforces the need for improved developer tools that integrate a wider variety of features, extend support to multiple programming languages, and help developers avoid a broader range of logging issues such as duplicate logging or readability problems. Similarly, unified platforms could be developed as IDE plug-ins to provide a one-stop solution for real-time logging-related assistance within the development environment, as highlighted in S13, S19, and S20.

 \item \textbf{Establishing Common Benchmarks and Logging Guidelines}

One important direction for future research is to develop standardized benchmarks and evaluation settings for log issue detection. Based on the findings of RQ3, it is clear that there is a lack of consistency in how tasks are defined, data is selected, and performance is measured. For example, many studies use logging statements written by developers as ground truth, but this is not standardized or verified across projects. Also, different studies use different sets of log statements, commits, or issue reports, which makes it difficult to compare results. The evaluation metrics also differ, as some studies report precision and recall, while others use accuracy or AUC. In addition, while some works focus on incorrect log levels and others on message inconsistency or duplication, there is no common task definition to guide these efforts. Because of all these differences, it becomes hard to evaluate and compare methods in a fair and reliable way. Therefore, future work should consider building shared datasets, unified task definitions, and common baseline models to improve consistency in this research area.

 Moreover, standardizing logging guidelines can help move toward more consistent research. A first step is to develop and promote community-accepted logging standards and checklists to guide practitioners and reduce inconsistencies in practice. Building on that, another step is to create datasets with labeled examples of problematic and well-formed log statements, which can support empirical evaluation  and allow researchers to assess and refine their techniques on shared benchmarks. Finally, integrating automated tooling that checks compliance with these guidelines can support their use in real projects.
 
\end{itemize}

\begin{tcolorbox}[colback=white, colframe=black, boxrule=0.5pt, sharp corners]
\textbf{Finding 12.} Future opportunities include enhancing log analysis techniques (e.g., LLMs), expanding support for multiple programming languages (e.g., Python), improving developer tools (e.g., to support multiple types of log smells), and establishing common benchmarks (e.g, datasets and evaluation metrics).
\end{tcolorbox}

\section{Threats to Validity}

In this section, we highlight possible validity concerns related to our study.

\subsection{Internal Validity }

In this study, we relied on the snowballing method rather than a traditional database search to identify the primary studies. This decision was based on our initial attempts with keyword-based searches, which produced many irrelevant results and made the screening process very time-consuming. The snowballing approach allowed us to start with a carefully chosen set of relevant papers and then expand it by following their references and citations. This helped us stay focused on studies closely related to our topic. Although there is still a risk of missing some important work, we tried to minimize this by testing different search strings to select strong initial papers, prioritizing those with high citation counts and clear relevance. We believe that any missing studies are likely few and would not significantly affect our findings.

Moreover, a number of the primary studies included in this review go beyond detection and also address repair. However, our analysis considered only the detection aspects, even when they were discussed briefly in some cases. This selective focus may introduce a threat to internal validity, as it might not fully capture the main objectives of those studies. As a result, there is a risk that some interpretations may overemphasize detection elements that were not the central contribution of the original work. To address this, we made sure to clearly state the scope of our review and carefully extracted only the relevant detection-related information from each study.

\subsection{Construct Validity}

Some of the categorizations used in answering the research questions were developed based on our interpretation of the studies, which may introduce subjectivity. While we aimed to ensure consistency by grounding categories in observed patterns and existing terminology, it is possible that alternative interpretations exist. We limited this threat by following a consistent process, refining the categories step by step, and referring back to the original studies to make sure our classifications matched their content.

Additionally, we mapped the logging issues to an existing taxonomy, which could introduce some interpretation bias. Sometimes, the way we assigned a study to a category might not fully reflect the original authors’ intention, especially if the study covered more than one area. To reduce this threat, we carefully studied the original taxonomy, examined the content and context of each primary study, and aimed to maintain consistency and clarity in our mapping decisions.

\subsection{External Validity}

Although some studies also address repair and a few are primarily repair-focused, this review concentrates solely on the detection of logging issues. As a result, the conclusions drawn are valid only within this specific scope and should not be generalized to repair-focused research.

\section{Conclusion}
This study provides a structured overview of existing research on detecting logging issues in software systems. By reviewing 21 primary studies, we examined how the problem is framed, what types of issues are addressed, which detection techniques are used, and how these techniques are evaluated. Our findings highlight several inconsistencies across the literature, ranging from variation in the number and types of issues addressed to the lack of shared benchmarks and evaluation practices. While the field has made significant progress, it remains fragmented, and the absence of standardized datasets, consistent validation methods, and unified task definitions continues to hinder comparison and reproducibility.

Despite these challenges, the increasing interest in automation, along with the potential of LLM techniques, presents new opportunities for advancing log issue detection. Future work should not only explore more intelligent detection approaches but also prioritize the creation of community-agreed benchmarks, shared tasks, and tooling support. In this way, the research community can move toward more unified, practical, and scalable solutions for improving log quality in modern software systems.

This review sheds light on how the research community has approached detection efforts so far. The findings make it clear that although progress has been made, there is still significant room for improvement and collaboration in this area.

\section*{Declaration of Generative AI and AI-assisted Technologies in the Writing Process}

During the preparation of this work the author(s) used ChatGPT in order to enhance the clarity, readability, and language of the text. After using these tools, the author(s) reviewed and edited the content as needed and take(s) full responsibility for the content of the published article.

\section*{Acknowledgment}
The authors would like to thank Ongoing Research Funding pro-
gram, (ORFFT-2025-048-2), King Saud University, Riyadh, Saudi Ara-
bia for financial support.

%Section text. See Subsection \ref{subsec1}.

%% Use \subsubsection, \paragraph, \subparagraph commands to 
%% start 3rd, 4th and 5th level sections.
%% Refer following link for more details.
%% https://en.wikibooks.org/wiki/LaTeX/Document_Structure#Sectioning_commands

%%\appendix
%%\section{Example Appendix Section}
%%\label{app1}

%%Appendix text.

%%       \cite{<label>} ==> [1]

%%
%%Example citation, See 




\bibliographystyle{IEEEtran} % or another style like plain, alpha, etc.
\bibliography{ref}    % assumes the file is references.bib

%% \begin{thebibliography}{00}

%% For numbered reference style
%% \bibitem{lamport94}
  %% Leslie Lamport,
  %% \textit{\LaTeX: a document preparation S02stem},
  %% Addison Wesley, Massachusetts,
  %% 2nd edition,
 %%  1994.

%% \end{thebibliography}
\end{document}